\documentclass[journal]{IEEEtran}
\usepackage{cite}
\usepackage{graphicx}
\usepackage{subfigure}
\usepackage{url}
\usepackage{epsfig}
\usepackage{amsmath}
\usepackage{amsfonts}
\usepackage{amssymb}
\usepackage{mathrsfs}
\usepackage[final,scrtime]{prelim2e}  
\usepackage{verbatim}
\usepackage{algorithm}
\usepackage{algorithmic}
\usepackage{ntheorem}
\usepackage{url}
\usepackage{color}
\usepackage{multirow}     
\usepackage{diagbox}    
\usepackage{array}      
\usepackage{caption}    
\usepackage{colortbl}
\usepackage{longtable}   
\usepackage{xcolor}
\usepackage{xpatch}
\usepackage{lipsum,mwe,cuted}
\usepackage{stfloats}
\usepackage{enumerate}  
\usepackage{amsmath,mathtools}
\usepackage{makecell}

\graphicspath{{./}}
\theorembodyfont{\upshape}
\theoremheaderfont{\rmfamily\itshape}
\theoremseparator{:}

\theoremstyle{remark}

\newcommand{\tabincell}[2]{\begin{tabular}{@{}#1@{}}#2\end{tabula}}

\theoremstyle{remark}

\makeatletter
\def\changeBibColor#1{%
\ifin@\color{red}\else\normalcolor\fi
}
\xpatchcmd\@bibitem
{\item}
{\changeBibColor{#1}\item}
{}{\fail}
\xpatchcmd\@lbibitem
{\item}
{\changeBibColor{#2}\item}
{}{\fail}
\makeatother

\def\mb{\mathbf}

\def\rm{\textrm}

\renewcommand{\vec}{\mb}    

\begin{document}
\title{Movable Antenna for Wireless Communications: Prototyping and Experimental Results}

\author{
Zhenjun Dong, Zhiwen Zhou, Zhiqiang Xiao, Chaoyue Zhang, Xinrui Li, Hongqi Min, Yong Zeng,~\IEEEmembership{Senior Member,~IEEE}, Shi Jin,~\IEEEmembership{Fellow,~IEEE}, and Rui Zhang,~\IEEEmembership{Fellow,~IEEE}


       \thanks{Z. Dong, Z. Zhou, Z. Xiao, C. Zhang, X. Li, H. Min, Y. Zeng, and S. Jin are with the National Mobile Communications Research Laboratory and Frontiers Science Center for Mobile Information Communication and Security, Southeast University, Nanjing 210096, China. Z. Xiao and Y. Zeng are also with the Purple Mountain Laboratories, Nanjing 211111, China (e-mail: \{zhenjun\_dong, zhiwen\_zhou, zhiqiang\_xiao, 220211013, 230218659, minhq, yong\_zeng, jinshi\}@seu.edu.cn). (Corresponding author: Yong Zeng.)}
       \thanks{R. Zhang is with the School of Science and Engineering, Shenzhen Research Institute of Big Data, The Chinese University of Hong Kong, Shenzhen, Guangdong 518172, China (e-mail: rzhang@cuhk.edu.cn). He is also with the Department of Electrical and Computer Engineering, National University of Singapore, Singapore 117583 (e-mail: elezhang@nus.edu.sg).}

}

\maketitle

\begin{abstract}
Movable antenna (MA), which can flexibly change the position of antenna in three-dimensional (3D) continuous space, is an emerging technology for achieving full spatial performance gains.
In this paper, a prototype of MA communication system with ultra-accurate movement control is presented to verify the performance gain of MA in practical environments.
The prototype utilizes the feedback control to ensure that each power measurement is performed after the MA moves to a designated position.
The system operates at 3.5 GHz or 27.5 GHz, where the MA moves along a one-dimensional horizontal line with a step size of $0.01\lambda$ and in a two-dimensional square region with a step size of $0.05\lambda$, respectively, with $\lambda$  denoting the signal wavelength.
The scenario with mixed line-of-sight (LoS) and non-LoS (NLoS) links is considered. 
Extensive experimental results are obtained with the designed prototype and compared with the simulation results, which validate the great potential of MA technology in improving wireless communication performance.
For example, the maximum variation of measured power reaches over 40 dB and 23 dB at 3.5 GHz and 27.5 GHz, respectively, thanks to the flexible antenna movement.
 In addition, experimental results indicate that the power gain of MA system relies on the estimated path state information (PSI), including the number of paths, their delays, elevation and azimuth angles of arrival (AoAs), as well as the power ratio of each path.


\end{abstract}

\section{Introduction}
The evolution of wireless communication systems is accompanied with the development of advanced antenna technology.
Starting from single-antenna communications in the early days, multiple-input multiple-output (MIMO) and massive MIMO have become the key physical layer technologies of the fourth-generation (4G) and fifth-generation (5G) mobile communication systems, respectively.
Throughout the process, the scale of antenna array is continuously expanded, which brings new design degrees-of-freedom (DoFs) in the spatial domain for improving the communication performance.
To support the ambitious goals of the sixth-generation (6G) mobile communication systems, massive MIMO is expected to be further developed towards extremely large-scale MIMO (XL-MIMO) \cite{zhang2020prospective,de2020non,lu2024tutorial}, by further increasing the array size by at least one order of magnitude.
However, simply increasing the antenna size will face some practical issues, like prohibitive hardware cost and energy consumption, due to the growing number of antenna elements and their associated radio frequency (RF) chains.
Therefore, in addition to traditional arrays with adjacent elements separated by half wavelength  \cite{zhang2020prospective,lu2021communicating,10562328}, various alternative antenna architectures were proposed \cite{li2024sparse,wang2023can,wang2024enhancing,zhou2024sparse,pal2010nested,vaidyanathan2010sparse,li2024multi,jeon2021mimo,zeng2024multi,7827017}, such as sparse MIMO \cite{li2024sparse,wang2023can,wang2024enhancing,zhou2024sparse,pal2010nested,vaidyanathan2010sparse}, modular MIMO \cite{li2024multi,jeon2021mimo,zeng2024multi}, and cell-free massive MIMO \cite{7827017}.

For the aforementioned antenna architectures, the antenna elements are deployed at fixed positions with either half-wavelength or wider antenna spacing.
However, due to the hard restriction of antenna aperture and spacing, the antenna array with fixed position cannot fully explore the spatial variation of wireless channels in a given area, thus the communication performance is poor in the case of deep spatial channel fading.
To overcome such limitations, a novel antenna architecture termed {\it movable antenna} (MA) was recently proposed\cite{zhu2023movable,10318061}.
The typical structure of MA consists of two modules, i.e., communication and antenna positioning modules.
The communication module is similar to that of fixed-position antenna (FPA) systems, where the MA is connected to the RF chain via a flexible cable to support the antenna movement.
Furthermore, the MA is fixed on mechanical sliding rails to support free movement in the three-dimensional (3D) space via the motors controlled by the antenna positioning module.
Note that the movement accuracy and response delay of MA depend on different hardware manufacturers, some of which can achieve tens of micro-meter level in terms of positioning accuracy, which corresponds to the sub-Terahertz wavelength.
Moreover, with the help of a servo motor, the MA at a given position can freely change its orientation, thereby offers three additional DoFs for the antenna adjustment.

Compared to the conventional FPA, the flexible adjustment of position and orientation of MA brings new design DoFs to alter the wireless channel conditions and improve the communication performance.
For example, for the basic scenario when signal arrives via two propagation paths with different angles-of-arrival (AoAs), if the receiving MA is located at the positions where the phases of the two signal components differ by integer multiples of $2\pi$,
the signal components will be constructively superimposed, leading to the maximum channel power gain \cite{zhu2023movable,10318061}, while they will be deconstructively superimposed if the MA is located at the positions where the phases differ by odd multiples of $\pi$.
This implies that the wireless channel power gain varies significantly with the antenna position.
For the conventional FPA, if the antenna array is located at the position corresponding to the deep channel fading, its communication performance will be poor.
By exploiting the MA technology, the complete spatial diversity gain can be exploited via dynamically moving the antenna elements to the desired positions.

Extensive efforts have been devoted to theoretically or numerically demonstrating the advantages of MA in wireless communications, such as signal power improvement, interference mitigation, flexible beamforming, and spatial multiplexing.
For example, in \cite{zhu2023movable}, the authors demonstrated the benefits of MA for improving the received signal-to-noise ratio (SNR) in both deterministic and stochastic channels.
In \cite{Zhu2023b,Xiao2023,Yang2024}, a multi-user MA communication system was considered, where the MA's position is optimized to help reshape the multi-user channels, thereby saving the total transmit power or improving the communication sum rate.
In \cite{Ma2024}, the authors utilized the MA in MIMO systems, where the MAs are deployed at both the transmitter (TX) and receiver (RX). It was demonstrated that by jointly optimizing the MAs' positions, the communication capacity of MIMO systems can be significantly improved.
In \cite{lu2024group}, the concept of group MA (GMA) was proposed, where a group of antenna elements can be moved collectively in a continuous manner, and a flexible sparse array architecture can be achieved through antenna selection.
In \cite{Ma2023} and \cite{Xiao2024}, the channel estimation of MA was investigated, where the channel response between transmit and receive antennas at any locations was reconstructed.
In addition, some other studies investigated the applications of MA in interference mitigation , e.g.,  \cite{Hu2024}. Furthermore, six-dimensional MA (6DMA) with both antenna position and rotation optimization has been investigated in \cite{Shao2024,shao20246D}.

Despite extensive theoretical studies on MA-based wireless communications, the promised performance gain of MA for wireless communication has not been validated in practical systems yet.
Some earlier works were reported to implement MA or similar concept for applications other than wireless communications, such as radar \cite{Zhuravlev2015}, navigation \cite{Li2022},
localization \cite{Sakamoto2011}, etc.
However, to the authors' best knowledge, the prototype and experimental results of MA for wireless communication have not been reported.
To fully harness the performance advantages of MA for future wireless communications, performance measurements that take into account practical issues, such as the propagation channel, the size of MAs, and movement accuracy, are necessary.
Therefore, this motivates our current work.

In this paper, we present the first of its kind prototype of MA for wireless communication and report extensive experimental results obtained with the prototype in practical environments.
Specifically, we consider a system where the TX and RX are equipped with an FPA and an MA, respectively.
The MA can be flexibly moved on a given two-dimensional (2D) plane.
The developed prototype is tested at 3.5 GHz or 27.5 GHz, corresponding to sub-6 GHz and millimeter wave (mmWave) frequency bands, respectively.
We use a slide track with a motor to practically implement the MA, with a positioning accuracy of 0.05 millimeters, which corresponds to less than 0.5\% wavelength even for 27.5 GHz.
The received signal power between the TX and the receiving region is measured.
Moreover, we estimate the number of received multipath components as well as the power delay spectrum (PDS) and power angular spectrum (PAS) via the received multi-frequency measurement data.
It was found that the experimental results are consistent with the simulation results based on the estimated multipath parameters.
Moreover, the maximum variation of signal power reaches over 40 dB by moving the MA within the wavelength range,  which verifies the enormous potential of MA in improving wireless communication performance.
Our main contributions are summarized as follows.
\begin{itemize}

\item First, a prototype of the MA communication system is developed, which consists of both the hardware platform and  software program. 
The prototype is implemented with a universal software radio peripheral (USRP), a host computer (HC), a transmitting FPA, and a receiving MA.
The system will initiate a measurement once feedback is received confirming that the slide track has moved to a  designated position.
The introduced feedback control ensures the successful implementation of the MA-based performance measurement.

\item
Then, the performance gain of MA wireless communications is verified by the developed prototype at both 3.5 GHz and 27.5 GHz. The experimental results are compared with the simulation results, which utilize the estimated channel parameters obtained from the multi-frequency measurement data. It is validated that MA communication can achieve significant performance gains over the conventional FPA system.

\item
In addition, experimental results indicate that the MA system relies on the estimated path state information (PSI)\cite{xiao2023exploiting,xiao2023integrated}, such as the number of multipath components, their delays, elevation AoAs, azimuth AoAs, and relative power ratios.
Therefore, we propose an efficient movement control scheme for MA, by combining channel sounding with coarse antenna movement and position refinement with light performance measurement. 
Specifically, the MA is first moved to an initial rough position, which can be determined based on the estimated PSI obtained from the channel sounding. Then, the MA is moved to a more accurate position according to a few performance measurements around the initial position.


\end{itemize}


The rest of the paper is organized as follows.
The theoretical model of MA communication system is presented  in Section II.
The developed prototype and experimental scenario are elaborated in Section III. Section IV introduces the experimental data processing for MA communication system. Experimental and simulation results are presented in Section V. Finally, we conclude this paper in Section VI.

{\it Notation}:
The transpose, Hermitian transpose, and complex conjugate operation are given by $(\cdot)^T$, $(\cdot)^H$, and $(\cdot)^{*}$, respectively.
$\mathbb{R}^2$ represents the 2D space of real numbers.
$j=\sqrt{-1}$ denotes the imaginary unit of complex numbers.
The distribution of a circularly symmetric complex Gaussian (CSCG) random variable with mean 0 and variance $\sigma^2$ is denoted by $\mathcal{CN}(0,\sigma^2)$, and
${\rm {round}}(\cdot)$ represents rounding to the nearest integer.

\section{system model}
\begin{figure}[htbp]
\centering
	{\includegraphics[width=1\linewidth]{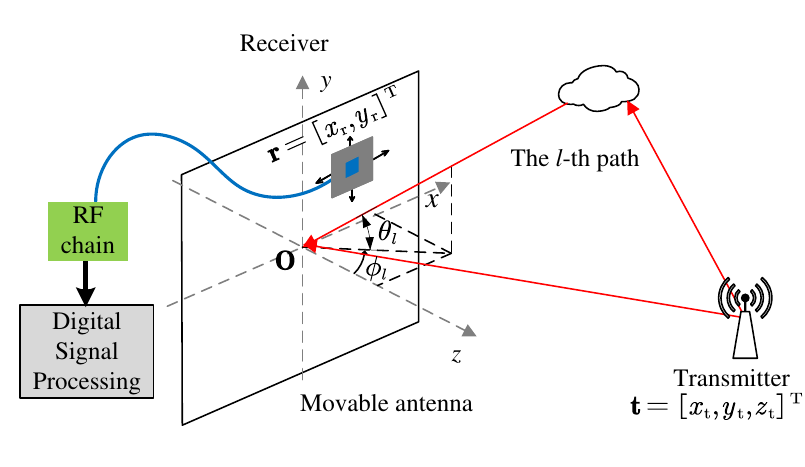}}
	\caption{Illustration of the MA communication system.}
	\label{coordinates}
\end{figure}
As illustrated in Fig.~\ref{coordinates}, we consider an MA wireless communication system, where the TX is equipped with an antenna at a fixed position, while the RX is equipped with an MA that can freely move on a given 2D plane. 
The center of the 2D plane for MA movement is set as the origin for the considered Cartesian coordinate system. Therefore, the coordinate of MA relative to the center of 2D plane is denoted by $\mathbf{r}=[x_\text{r},y_\text{r}]^T\in\mathcal{C}_\text{r}$, where $\mathcal{C}_\text{r}\subset\mathbb{R}^{2}$ denotes the area where the MA can move. Furthermore, denote by $\mathbf{t}=[x_\text{t},y_\text{t},z_\text{t}]^T$ the coordinate of the transmit antenna.
Therefore, for the considered MA wireless communication system, the channel response depends on the positions of TX and RX.
Denote by $h(\mathbf{t},\mathbf{r})$ the propagation channel between the TX and RX, which will be specified later in \eqref{reference model} and \eqref{channel response}.
Thus, the received signal can be expressed as
\begin{equation}\label{channel_h}
    y(\mathbf{t},\mathbf{r}) = h(\mathbf{t},\mathbf{r})\sqrt{p_t}s + z, \forall\mathbf{r}\in\mathcal{C}_\text{r},
\end{equation}
where $p_t$ denotes the transmit power, $s$ represents the transmit signal with the normalized power, and $z\sim\mathcal{CN}(0,\sigma^2)$ is the additive white Gaussian noise with power $\sigma^2$.
Thus, the received signal power without considering noise is
\begin{equation}\label{SNR}
    p_r(\mathbf{t},\mathbf{r})=|h(\mathbf{t},\mathbf{r})|^2p_t, \forall\mathbf{r}\in\mathcal{C}_\text{r}.
\end{equation}

Note that when the position of the transmit antenna $\mathbf{t}$ is fixed, the received signal power depends on the channel response $h(\mathbf{t},\mathbf{r})$ with respect to (w.r.t.) the position of the MA, i.e., $\mathbf{r}$ in the RX region $\mathcal{C}_r$.
To obtain the channel response, a far-field assumption is made in this paper and a field-response based model is adopted \cite{10437024}.
Specifically, denote by $L$ the number of multipath components and $\mathbf{r}_0=[0,0]^T$ the reference position at the RX.
The propagation channel between the TX and RX at position $\mathbf{r}_0$ is
\begin{equation}\label{reference model}
    h(\mathbf{t},\mathbf{r}_0) = \sqrt{\beta}\mathbf{1}_{L}^H\mathbf{b}(\mathbf{t},\mathbf{r}_0),
\end{equation}
where $\beta$ denotes the large-scale channel gain, $\mathbf{b}(\mathbf{t},\mathbf{r}_0)=\left[b_1(\mathbf{t},\mathbf{r}_0),\cdots,b_{L}(\mathbf{t},\mathbf{r}_0)\right]^T\in\mathbb{C}^{L\times1}$ is the channel coefficient vector, and $b_l(\mathbf{t},\mathbf{r}_0)$ is the channel coefficient of the $l$-th path, which is given by
\begin{equation}\label{b_l}
   b_l(\mathbf{t},\mathbf{r}_0)=\alpha_l(\mathbf{t},\mathbf{r}_0)e^{-j2\pi f_c\tau_l(\mathbf{t},\mathbf{r}_0)},
\end{equation}
where $\alpha_l(\mathbf{t},\mathbf{r}_0)$ is the small-scale amplitude of the $l$-th path satisfying $\sum_{l=1}^{L}\alpha_l^2(\mathbf{t},\mathbf{r}_0)=1$;  $f_c$ represents the carrier frequency; and $\tau_l(\mathbf{t},\mathbf{r}_0)$ denotes the propagation delay of the $l$-th path.
Note that the small-scale amplitude  $\alpha_l(\mathbf{t},\mathbf{r}_0)$ and propagation delay $\tau_l(\mathbf{t},\mathbf{r}_0)$ depend on the relative position of the TX and RX.
Denote by $\theta_l\in\left[-\frac{\pi}{2},\frac{\pi}{2}\right]$ and $\phi_l\in\left[-\frac{\pi}{2},\frac{\pi}{2}\right]$ the elevation and azimuth AoA of the $l$-th path, respectively.
For the RX at an arbitrary position $\mathbf{r}\in\mathcal{C}_\text{r}$, the variation of the propagation distance of the $l$-th path w.r.t. that at $\mathbf{r}_0$ is given by $d_l(\mathbf{r})=x_\text{r}\cos\theta_l\sin\phi_l+y_\text{r}\sin\theta_l$.
This leads to a phase variation of $\frac{2\pi}{\lambda}d_l(\mathbf{r})$
at $\vec{r}$ w.r.t. that at $\mathbf{r}_0$ for the $l$-th path, where $\lambda$ is the signal wavelength.
Therefore, to account for the phase variations of all $L$ paths, a field-response vector (FRV) $\mathbf{f}(\mathbf{r})\in\mathbb{C}^{L\times1}$ at $\vec{r}$ is defined as
\begin{equation}\label{f_r}
    \mathbf{f}(\mathbf{r}) = \left[e^{j\frac{2\pi}{\lambda}d_1(\mathbf{r})},\cdots,e^{j\frac{2\pi}{\lambda}d_L(\mathbf{r})}\right]^T , \forall\mathbf{r}\in\mathcal{C}_\text{r}.
\end{equation}

Next, we assume that the MA is moved in a small (wavelength-scaled) region and thus, the variation of $\alpha_l(\mathbf{t},\mathbf{r}_0)$ over $\vec{r}$ is neglected, i.e., $\alpha_l(\mathbf{t},\mathbf{r})\approx\alpha_l(\mathbf{t},\mathbf{r}_0)$, $\forall l, \mathbf{r}\in\mathcal{C}_\text{r}$.
As a result, the propagation channel between the TX at a fixed position $\vec{t}$ and RX at an arbitrary position $\mathbf{r}\in\mathcal{C}_\text{r}$ is given by
\begin{equation}\label{channel response}
    h(\mathbf{r}) =\sqrt{\beta} \mathbf{f}(\mathbf{r})^H\mathbf{b}, \forall\mathbf{r}\in\mathcal{C}_\text{r},
\end{equation}
where $\vec{b}=\vec{b}(\mathbf{t},\mathbf{r}_0)=[b_l(\mathbf{t},\mathbf{r}_0)]_{1\leq l\leq L}$, where
$b_l(\mathbf{t},\mathbf{r}_0)=\alpha_le^{-j2\pi f_c\tau_l}$, with
 $\alpha_l=\alpha_l(\mathbf{t},\mathbf{r}_0)$ and $\tau_l=\tau_l(\mathbf{t},\mathbf{r}_0)$ denoting the small-scale amplitude and propagation delay of the $l$-th path
 corresponding to the position pair ($\vec{t},\vec{r}_0$).
By substituting \eqref{b_l}, \eqref{f_r} and \eqref{channel response} into \eqref{SNR}, the signal power is rewritten as
\begin{equation}\label{received_power}
\begin{aligned}
    p_r(\mathbf{r}) &= \left|\sqrt{\beta}\mathbf{f}(\mathbf{r})^H\mathbf{b}\right|^2p_t\\
                       &=  p_t\beta\left|\sum\limits_{l=1}^{L}\alpha_l\exp\left(-j2\pi\left(\frac{d_l(\mathbf{r})}{\lambda}+f_c\tau_l\right)\right)\right|^2\\
                       &= G g(\mathbf{r}),
\end{aligned}
\end{equation}
where $G=p_t\beta$ is the local-average received power, and $g(\vec{r})\triangleq\left|\sum\limits_{l=1}^{L}\alpha_l\exp\left(-j2\pi\left(\frac{d_l(\vec{r})}{\lambda}+f_c\tau_l\right)\right)\right|^2$ is the small-scale channel power gain of the MA at position $\vec{r}$.
Note that by varying the position of the MA, it is expected that the MA can achieve higher channel gain if the multipath components are superimposed constructively.
However, such a potential performance gain has not been verified in practice yet, especially for the MA driven by motor.
To fill this gap, we develop a prototype of MA communication system and use it to measure the received signal power by changing the MA's position, so as to verify its performance gain in practical environments.

\section{Prototyping of MA Wireless Communication}
\subsection{Prototype Architecture} \label{Prototype architecture}
\begin{figure}[!hb]
\centering
	{\includegraphics[width=0.95\linewidth]{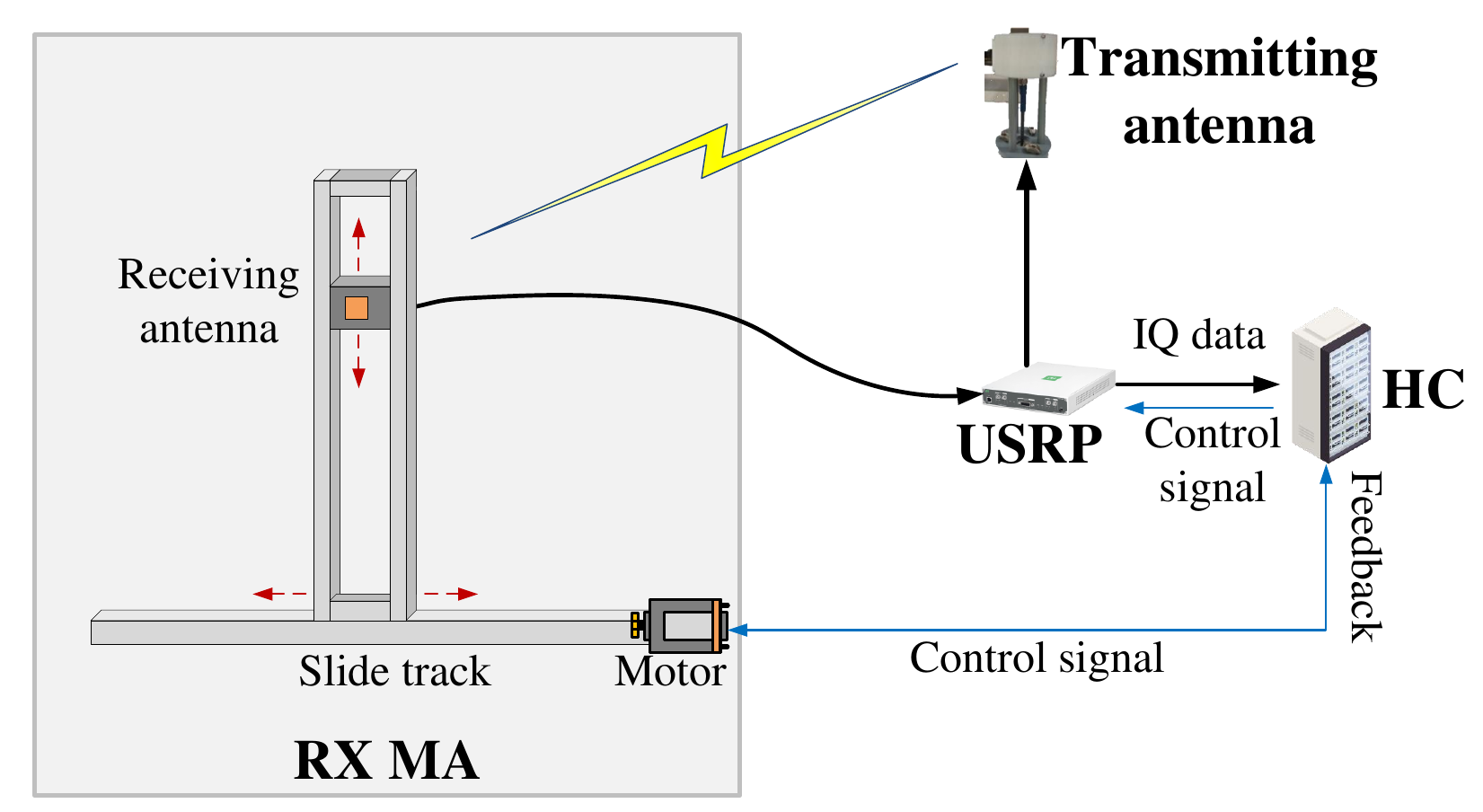}}
	\caption{Architecture of the developed prototype for MA wireless communication.}
	\label{system_structure}
\end{figure}

In order to verify the communication performance of MA technology, an experimental prototype of MA wireless communication system is developed.
Specifically, as shown in \mbox{Fig.~\ref{system_structure}}, the prototype architecture consists of a USRP, an HC, a fixed transmitting antenna, and an RX MA.
The USRP serves as the basic component in the MA prototype, which handles both baseband signal processing and intermediate frequency (IF) transceiving.
The HC is utilized to realize the movement control of slide tracks, experimental measurement control, and processing of experimental in-phase and quadrature (IQ) baseband data.
In addition, the RX MA consists of a receiving antenna and a high-precision slide track, which can move in both vertical and horizontal directions.
Furthermore, different communication protocols are applied between different devices. Specifically, the transmitting and receiving antennas are connected to USRP via soft RF cables, while the HC communicates with the slide track via a controller area network (CAN) bus. Additionally, USRP sends IQ data to HC via an MXI-Express cable, and HC sends the measurement control signals to USRP via an Ethernet cable.

In order to study the communication performance of MA system, the received power of the RX MA is measured in the experiment.
In addition, channel sounding is used to estimate the parameters of the multipath components to further verify the practical measurement results.
To ensure cross-checking, performance measurement and channel sounding are designed as two separate procedures.
A single-frequency sinusoidal wave signal is used for performance measurement, for that the received signal power can be easily obtained by using fast Fourier transform (FFT), which is less vulnerable to the influence of noise and interference, while the signal for channel sounding is orthogonal frequency division multiplexing (OFDM) modulated  with a bandwidth of 400 MHz.

\begin{figure}[htbp]
\centering
\subfigure[]
{
    \begin{minipage}[b]{.5\linewidth}
        \centering
         \includegraphics[height = 3cm]{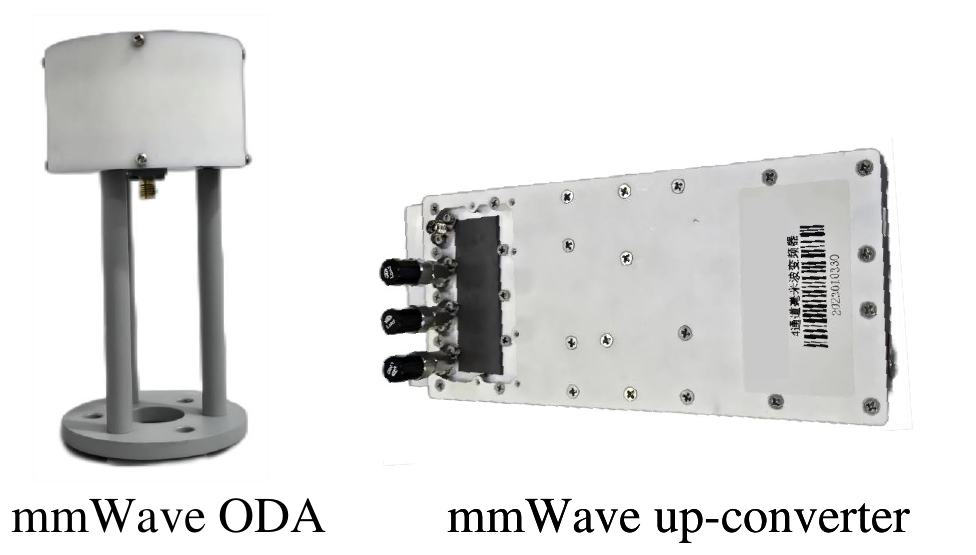} \\
        \hspace{1cm}
        \includegraphics[height = 2cm]{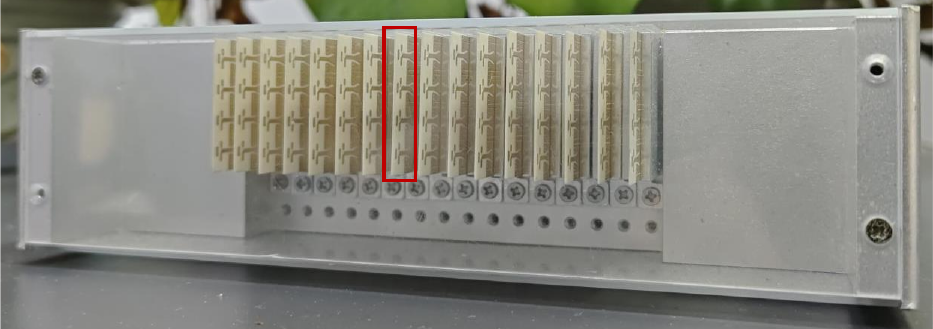}
    \end{minipage}
}
\hspace{1cm}
\subfigure[]
{
    \begin{minipage}[b]{.3\linewidth}
        \centering
        \includegraphics[height = 5cm]{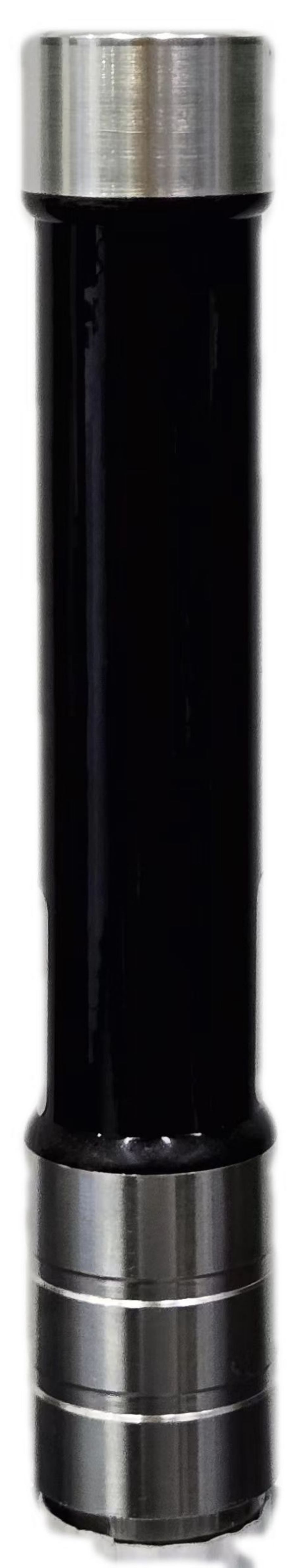}
    \end{minipage}
}
\caption{Transmitting and receiving antennas at different carrier frequencies (a) transmitting antenna (top) and receiving antenna (bottom) at 27.5 GHz, (b) transmitting/receiving antenna at 3.5 GHz.}
\label{Ant}
\end{figure}

The experimental measurements consider carrier frequencies of 3.5 GHz and 27.5 GHz.
For different carrier frequencies, the TX and RX need to configure different types of antennas.
Specifically, in the case of 27.5 GHz, the transmitting antenna consists of a mmWave omnidirectional antenna (ODA) and a mmWave up-converter, as illustrated at the top of \mbox{Fig.~\ref{Ant}} (a);
and only the eighth antenna element of the mmWave phased array (mmPSA) is enabled as the receiving antenna at the bottom of \mbox{Fig.~\ref{Ant}} (a).
Since the mmPSA has its own down-converter, the combination of the mmWave ODA and mmWave up-converter is equivalently regarded as the transmitting antenna.
Besides, for 3.5 GHz, the sub-6 GHz ODAs serve as the transmitting and receiving antennas, as shown in \mbox{Fig.~\ref{Ant}} (b).



\subsection{Hardware Equipment}\label{hard_equp}
The main hardware components of the developed prototype for MA communication system include USRP, HC, sub-6 GHz ODA, mmWave ODA, mmPSA, mmWave up-converter, and high-precision slide track. The specifications and configurations of the main hardware components are given as follows.

\subsubsection{USRP}
The NI USRP-X410 is an advanced stand-alone software defined radio (SDR) device from National Instruments, which incorporates an RF system on chip (RFSoC) for rapid measurement and prototyping of high-performance wireless communication systems. It provides a real-time bandwidth of up to 400 MHz with two daughter boards covering a frequency range from 1 MHz to 7.2 GHz to perform analog up/down conversion. 
Paired with NI Labview software, the USRP-X410 can be employed as a versatile sub-6 GHz transceiver.

\subsubsection{mmPSA} The mmPSA TR16-1909 is a 16-element mmWave uniform linear phased array. Each antenna element essentially corresponds a small 4-element vertical sub-array with the same phase. The phase and on-off state of each element can be controlled individually to provide maximum flexibility. The device is also equipped with an integrated mmWave up/down-converter, which can up convert IF signals in the range of 1.8-3.8 GHz to mmWave signals in the range of 27-29 GHz, or down convert mmWave signals to IF signals. 
In the developed prototype, the eighth element is activated to enable the mmPSA to act as a single mmWave antenna.

\subsubsection{mmWave up-converter} 
The mmFE-28-4CM mmWave up-converter is a 27-29 GHz time division duplexing (TDD) 4-channel mmWave transceiver, which integrates mmWave low-noise amplifiers (LNAs) and mmWave power amplifiers (PAs). It has excellent noise characteristics and transmission signal accuracy, making it suitable for applications such as mmWave communication testing, instrument measurement, and production line testing. In the developed prototype, one channel of the mmWave up-converter is utilized to up convert the IF signal from USRP to the mmWave signal, which is then sent to the mmWave ODA for transmission.

\subsubsection{High-precision slide track} The high-precision slide track can move in both vertical and horizontal directions, with a total length of 1.5 m for each axis and a positioning accuracy of 0.05 mm.

The corresponding parameters for the developed prototype for different carrier frequencies are specified in \mbox{Table~\ref{Parameters0}}.

\begin{table}[htbp]
    \caption{Parameters for the developed prototype of MA wireless communication system.}
    \centering
    \begin{tabular}{|c|cccc|}
    \hline
    Polarization                 & \multicolumn{4}{c|}{Vertical}       \\ \hline

    \makecell{Signal for \\power measurement} & \multicolumn{4}{c|}{Single-frequency sinusoidal wave}                                                                                                                                                                                                                                                             \\ \hline
    \makecell{Signal for \\channel sounding}  & \multicolumn{4}{c|}{OFDM}                                                                                                                                                                                                                                                             \\ \hline
    \makecell{OFDM \\sub-carrier spacing}  & \multicolumn{4}{c|}{120 kHz}                                                                                                                                                                                                                                                             \\ \hline
    \makecell{Number of \\OFDM sub-carriers}  & \multicolumn{4}{c|}{3168}                                                                                                      \\ \hline
    \makecell{Number of\\ OFDM symbols}  & \multicolumn{4}{c|}{100}                                                                                                      \\ \hline
    Slide track length  & \multicolumn{4}{c|}{1.5 m}    \\ \hline
    \makecell{Positioning accuracy\\ of slide track}                & \multicolumn{4}{c|}{0.05 mm}
                                            \\ \hline
     Center frequency             & \multicolumn{2}{c|}{3.5 GHz}                                                                                                                       & \multicolumn{2}{c|}{27.5 GHz}                                                                                                             \\ \hline
    Antenna type                 & \multicolumn{1}{c|}{\begin{tabular}[c]{@{}c@{}}ODA\\ (RX)\end{tabular}} & \multicolumn{1}{c|}{\begin{tabular}[c]{@{}c@{}}ODA\\ (TX)\end{tabular}} & \multicolumn{1}{c|}{\begin{tabular}[c]{@{}c@{}}Element of\\  mmPSA \\ (RX)\end{tabular}} & \begin{tabular}[c]{@{}c@{}}ODA\\ (TX)\end{tabular} \\ \hline
       \makecell{Move range/area for \\ power measurement  }        & \multicolumn{2}{c|}{1D: 500 mm} & \multicolumn{2}{c|}{2D: 50 mm$\times$ 50 mm}\\ \hline
       \makecell{Move step size \\for power measurement}               & \multicolumn{2}{c|}{1 mm} & \multicolumn{2}{c|}{0.5 mm}
                                            \\ \hline
       \makecell{Move area \\for channel sounding}         & \multicolumn{2}{c|}{500 mm $\times$ 500 mm} & \multicolumn{2}{c|}{50 mm$\times$ 50 mm}\\ \hline
       \makecell{Move step size \\for channel sounding }              & \multicolumn{2}{c|}{5 mm} & \multicolumn{2}{c|}{1 mm}
                                            \\ \hline
    TX height                    & \multicolumn{4}{c|}{1.3 m}
                                            \\ \hline
    \end{tabular}
    \label{Parameters0}
    \end{table}


\subsection{Program Design}

\begin{figure}[htbp]
\centering
\includegraphics[width=1\columnwidth]{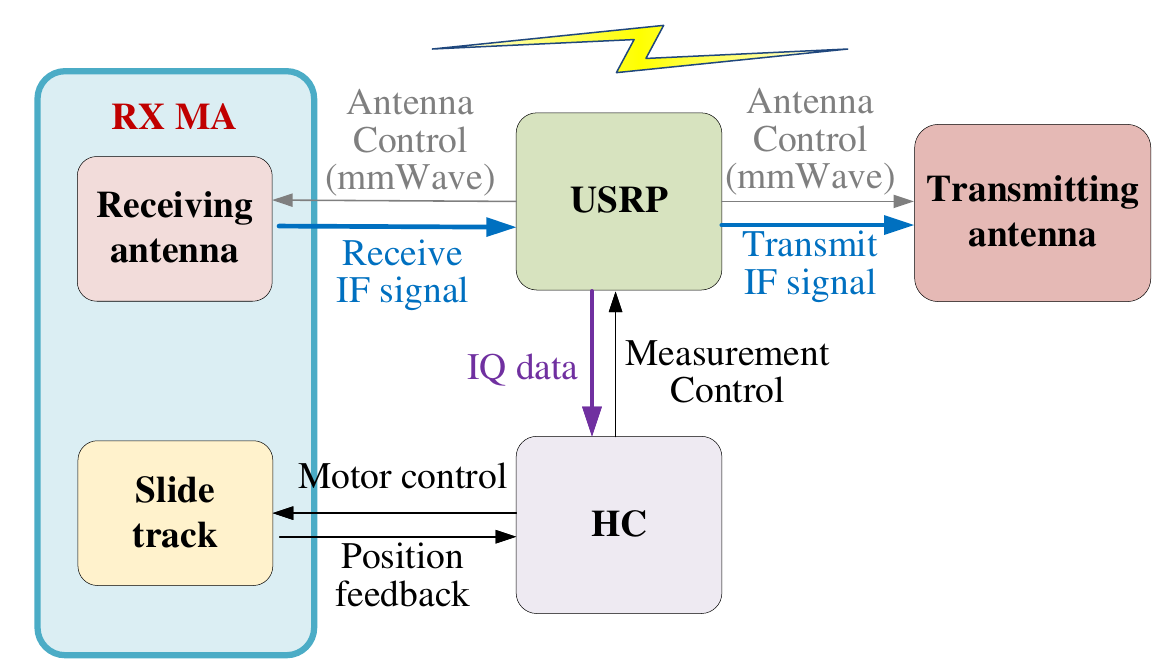}
\caption{Program design of the MA communication system prototype.}
\label{fig:prog_design}
\end{figure}

Customized program is designed to perform experimental measurements at different carrier frequencies for the MA communication system prototype, as shown in \mbox{Fig.~\ref{fig:prog_design}}, which is based on the hardware devices introduced in Section \ref{hard_equp}.
Different from the sub-6 GHz scenario, the mmWave measurement requires additional initialization of the transmitting and receiving antennas before measurement.
During the initialization process, the USRP sends antenna control commands to the transmitting and receiving antennas, to set the transceiver mode and power gain of the mmWave converters.

During the measurement process, the HC first sends a motor control command to the slide track to control its precise movement to the designated position.
After the MA moves to a designated position, the slide track will send position feedback to the HC.
The introduced feedback control ensures that each measurement is performed after the precise movement of MA.
Next, after receiving the position feedback, the HC will send a measurement control signal to trigger USRP for measurement.
Subsequently, the USRP modulates a pre-designed baseband sequence into an IF signal, and then sends it to the transmitting antenna.
For the sub-6 GHz scenario, the IF signal will be transmitted directly. Otherwise, in the mmWave scenario, the IF signal will be further up-converted to mmWave signal at the transmitting antenna and then be transmitted.
 Correspondingly, at the receiving antenna, the received signal is sent back to USRP directly in the sub-6 GHz scenario, or is first down-converted back to an IF signal and then sent back to USRP in the mmWave scenario.
Finally, the USRP demodulates the received IF signal into baseband IQ signal and then sends it to the HC, which stores the measurement data indexed by position for further signal processing.

\subsection{Measurement Scenario}
\begin{figure}[htbp]
\centering
\subfigure[Practical measurement scenario]{
	\includegraphics[width=0.97\linewidth]{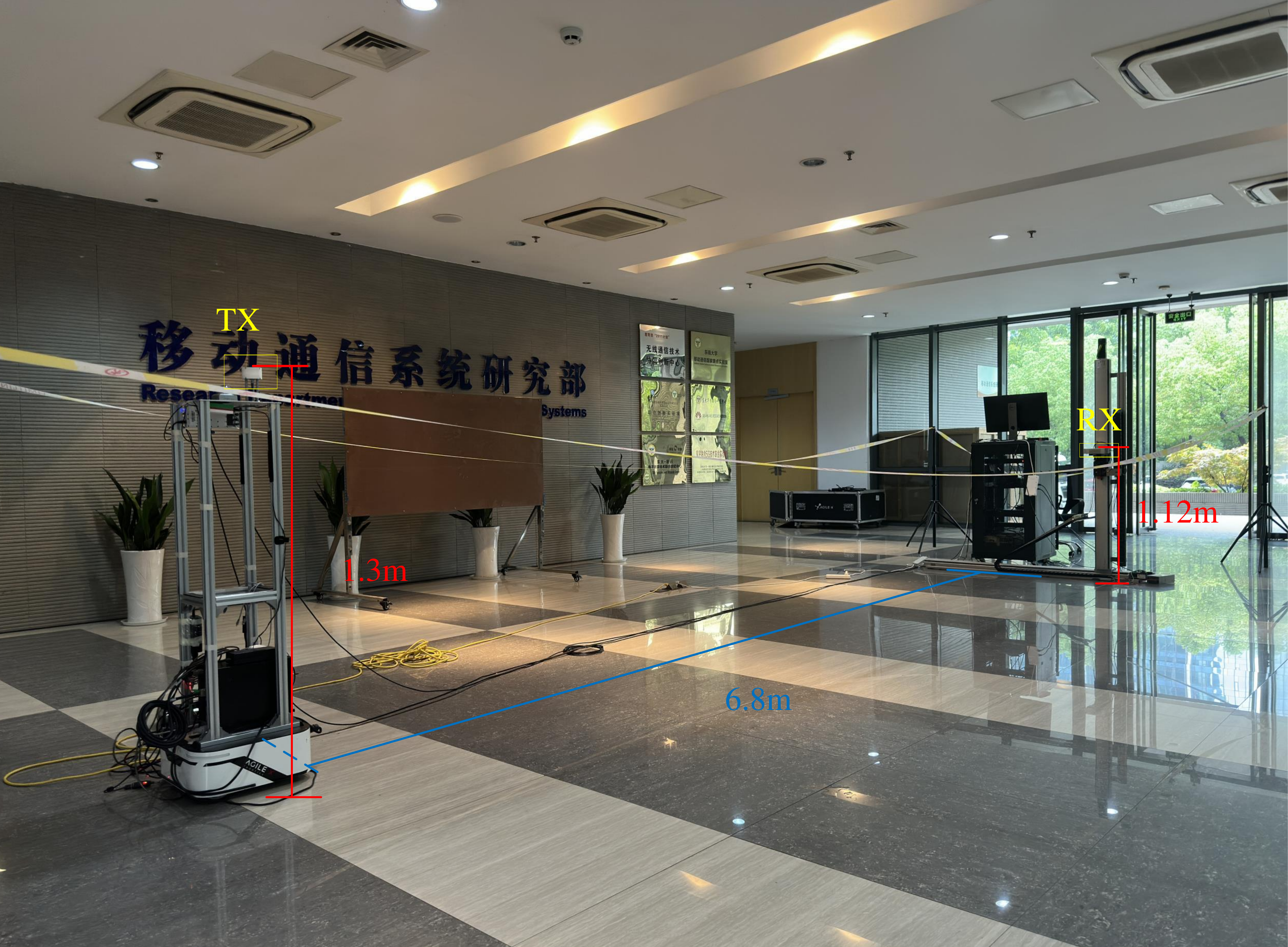}
        \label{actual} }
\subfigure[Experiment scenario setup]{
	\includegraphics[width=0.97\linewidth]{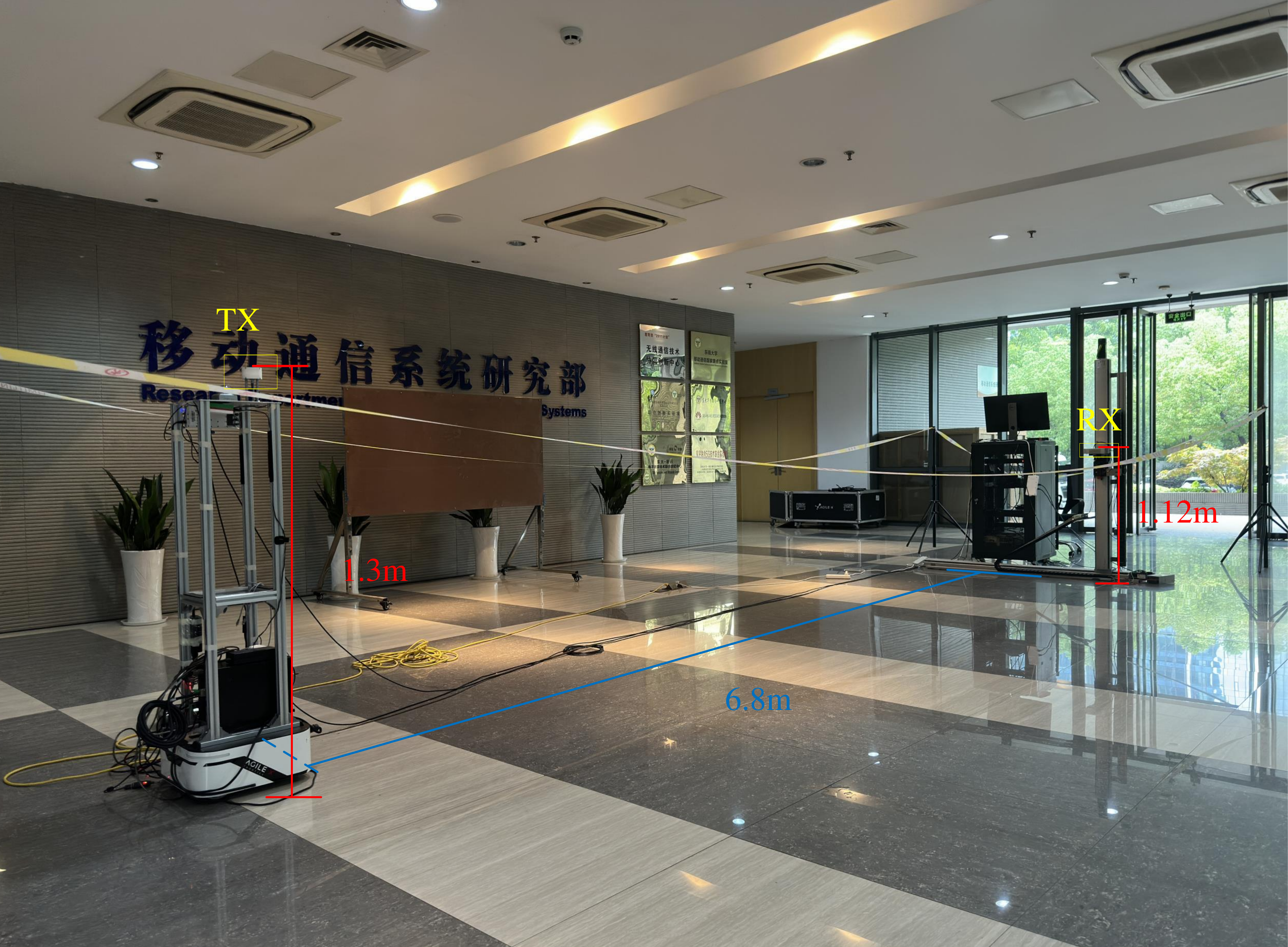}
        \label{setup}}
\caption{Measurement scenario of MA communication system prototype.}
\label{fig:7}
\end{figure}

The actual experiment measurement is conducted in a hall as shown in \mbox{Fig.~\ref{actual}}, whose size is about $21.7 \, \text{m}\times 8.8  \, \text{m}\times 3  \, \text{m}$.
The experimental hall is composed of flower pots, cement walls, glass doors, windows, copper plates on the walls, and a $2 \, \text{m} \times 1 \, \text{m}$ metal plate.
The measurement scenario involves the TX located at two positions, denoted by TX1 and TX2, respectively. The RX MA is set to move along a one-dimensional (1D) axis and in a 2D plane in sub-6 GHz and mmWave measurements, respectively.
The specific experiment scenario setup is illustrated in \mbox{Fig.~\ref{setup}}.
The center of the slide track is set as the origin, and the coordinates of the TX1 and TX2 are set to $[0,1.3,6.8]^T$ m and $[-0.8,1.3,6.8]^T$ m, respectively, with a height of 1.3 m. The coordinate of the center of the metal plate is $[3.45,1.2,3.8]^T$ m, whose bottom is 0.7 m above the ground.
In addition, the RX MA starts from the position $[-0.75,1.12,0]^T$ m, which moves to the right along the x-axis for sub-6 GHz, and moves to the right along the x-axis and downwards along the y-axis for mmWave.

During each measurement process, the RX MA moves by a preset step size, and the measurement results labeled with position are recorded. Specifically, in the case of performance measurement at sub-6 GHz, the movement range and step size of the RX MA are set to 500 mm and 1 mm, respectively, while in the mmWave scenario, they are set to 50 mm $\times$ 50 mm and 0.5 mm, respectively.
Since both azimuth and elevation AoAs need to be estimated, it is necessary to perform channel sounding in the 2D planar region.
Furthermore, since the PSI remains constant within the measurement region of the wavelength range, high-precision movement step sizes are not required in channel sounding.
Therefore, for channel sounding at sub-6 GHz, the MA is set to move in a 2D square region of 500 mm $\times$ 500 mm with movement step sizes of 5 mm in both horizontal and vertical directions, where the center of the square region coincides with that of the performance measurement region.
Besides, in the scenario of mmWave, the MA is set to move in the same region for channel sounding as that for performance measurement with step sizes of 1 mm in both horizontal and vertical directions.
The detailed configurations of measurement scenario are specified in \mbox{Table~\ref{Parameters0}}.

\section{DATA PROCESSING}

In this section, the data processing for measured power and estimation of channel parameters will be discussed, which are
obtained through performance measurement and channel sounding in MA communication system, respectively.
The power measurement with noise $\hat p_r({\bf r})$ is obtained by the FFT scheme, while the channel parameters $\{\hat \theta_l,\hat\phi_l,\hat \alpha_l,\hat \tau_l\}_{l=1}^{\hat{L}}$, PAS, and PDS are estimated based on the methods of OFDM and signal correlation.


\subsection{Data Processing for Power Measurement}
\begin{figure}[htbp]
\centering
\includegraphics[width=1\columnwidth]{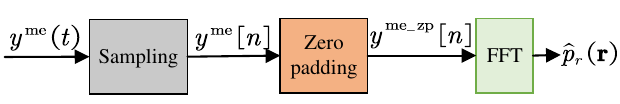}
\caption{The procedure of data processing for power measurement in the MA communication system prototype.}
\label{signalprocessing1}
\end{figure}
The procedure of data processing for power measurement in the MA communication system prototype is illustrated in Fig. \ref{signalprocessing1}. Since the transmitting baseband signal is set to the complex sinusoidal signal of frequency $f_0$, i.e., $s(t)=e^{j2\pi f_0t}$, the measured baseband analog signal at the RX in (\ref{channel_h}) in the time domain is expressed as
\begin{equation}\label{EQU-1} \vspace{-3pt}
y^{\text{me}}(t)=h({\bf r})\sqrt{p_t}e^{j2\pi f_0t}+z(t), {\bf r}\in {\mathcal C}_r,
\end{equation}
where $z(t)$ represents the band-limited complex Gaussian noise with power $\sigma^2$ and bandwidth $B$.
It is noted that we have omitted the antenna position notations $\bf t$ and $\bf r$ in the received signal $y^{\text{me}}(t)$ for convenient description.
By setting the sampling interval to $T=\frac{1}{B}$, the
sampled signal $y^{\text{me}}[n]=y^{\text{me}}(nT)$ is rewritten as
\begin{equation}\label{EQU-2} \vspace{-3pt}
y^{\text{me}}[n]=\left(h({\bf r})\sqrt{p_t}e^{j2\pi f_0nT}+z(nT)\right){\rm {rect}}_N[n], {\bf r}\in {\mathcal C}_r,
\end{equation}
where ${\rm {rect}}_N[n]$ is defined as
\begin{equation}\label{EQU-3}
{\rm {rect}}_N[n]=
\left\{
\begin{aligned}
&1,\quad n=0,\dots,N-1\\
&0,\quad\text{otherwise}\\
\end{aligned}
\right.
\end{equation}
with $N$ being the number of sampling points.
Next, zero padding is implemented to increase the number of sampling points in FFT, then we have
\begin{equation}\label{y0}
y^{\text{me}\_\text{zp}}[n]=
\left\{
\begin{aligned}
y^{\text{me}}[n],\quad &0\leq n\leq N-1\\
0,\quad& N\leq n\leq N_s-1\\
\end{aligned}
\right.
\end{equation}
where $N_s$ represents the size of FFT. Then, by performing FFT, we have
\begin{equation}\label{EQU-7}
\begin{split}
Y_{{\rm {FFT}}}[k]&=\text{FFT}\{y^{\text{me}\_\text{zp}}[n]\}=\sum_{n=0}^{N_s-1}y^{\text{me}\_\text{zp}}[n]e^{-j2\pi k \frac{n}{N_s}}\\
&=N{ h}({\bf r})\sqrt{p_t}{{\rm{sinc}}\bigg(\frac{N}{N_s}k-NTf_0\bigg)}+Z\bigg(\frac{k}{N_sT}\bigg),
\end{split}
\end{equation}
where $Z\left(\frac{k}{N_sT}\right)$ denotes the FFT of $z(nT){\rm {rect}}_N[n]$ after zero padding with $0\leq k\leq N_s-1$.
As such, the maximum value $\left|Y_{{\rm {FFT}}}[\hat k]\right|^2$ can be obtained with $\hat k={\rm {round}}(N_s T f_0)$.
In this case, the measured power with noise $\hat p_r({\bf r})$ can be obtained as
\begin{equation}\label{EQU-8} \vspace{-3pt}
{\hat p_r}({\bf r})=\frac{\left|Y_{{\rm {FFT}}}[\hat k]\right|^2}{N^2} .
\end{equation}\par
According to the measured power ${\hat p_r}({\bf r})$, we can describe the communication performance of MA system in the next section.\par

\subsection{Estimation of Channel Parameters}\label{Estimation of Channel Parameters}

\begin{figure}[htbp]
\centering
	\includegraphics[width=1\columnwidth]{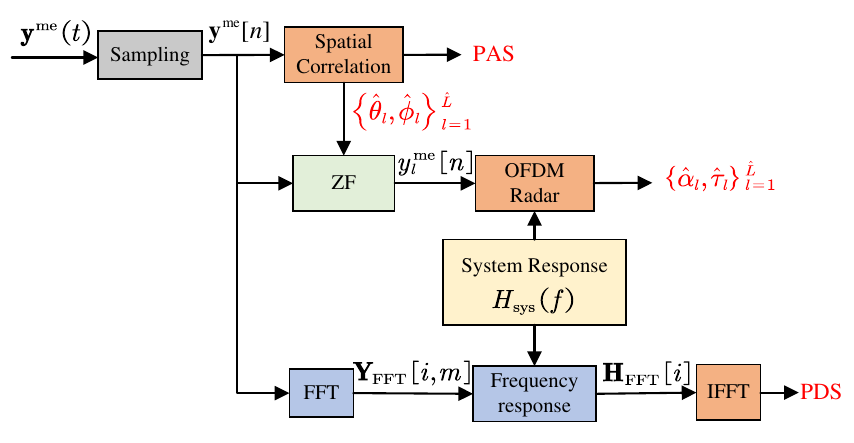}
\caption{The procedure of data processing for channel estimation in the MA communication system prototype.}
\label{signalprocessing2}
\end{figure}
In Fig. \ref{signalprocessing2}, the procedure of data processing for channel estimation in the MA communication system prototype is illustrated.
The number of multipath components $L$, elevation and azimuth AoAs $\{\theta_l,\phi_l\}_{l=1}^L$, and the PAS are obtained through the signal correlation method, while the small-scale amplitudes and delays of multipath components $\{\alpha_l,\tau_l\}_l^L$ are estimated by utilizing the OFDM radar method \cite{2014ofdmradar}. In addition, the PDS can be obtained by using the inverse FFT (IFFT) on the channel transfer function.
First, the transmitted OFDM signal with unit power is given by
\begin{equation}\label{EQU-90}
s(t)=\sum_{m=0}^{M-1}\sum_{i=0}^{I-1}b_{i,m}e^{j2\pi i \Delta f (t-mT_o-T_c)}{\rm{ rect}}\left(\frac{t-mT_o}{T_o}\right),
\end{equation}
where $M$ is the number of OFDM symbols; $I$ is the number of sub-carriers;  $\Delta f$ is the sub-carrier spacing;
$T_o$ and $T_c$ denote the duration of the OFDM symbol and cyclic prefix, respectively; and $b_{i,m}$ is the modulation symbol with ${\mathbb E}\{|b_{i,m}|^2\}=\frac{1}{I}$. Thus, based on (\ref{channel_h}), the received baseband
signal ${\bf y}^{\text{me}}(t)\in \mathbb{C}^{Q\times 1}$ for all the $Q$ MA possible locations is given by
\begin{equation}\label{EQU-90}
{\bf y}^{\text{me}}(t)=\sum_{l=1}^L{\bf h}_l\sqrt{p_t}s(t-\tau_{l})+{\bf z}(t),
\end{equation}
where
${\bf h}_l=\sqrt{\beta}\alpha_le^{-j2\pi f_c\tau_l}{\bf f}(\theta_l,\phi_l)\in \mathbb{C}^{Q\times 1}$ denotes the channel coefficient vector of path $l$, 
with ${\bf f}(\theta_l,\phi_l)\in \mathbb{C}^{Q\times 1}$ being the array response vector for all $Q$ MA locations w.r.t. path $l$, which is expressed as
\begin{equation}\label{array response}
{\bf f}(\theta_l,\phi_l)=\left[e^{-j\frac{2\pi}{\lambda}d_l({\bf r}_q)}\right]_{1\leq q\leq Q},
\end{equation}
with $d_l(\mathbf{r}_q)=x_{\text{r}_q}\cos\theta_l\sin\phi_l+y_{\text{r}_q}\sin\theta_l$ and $\mathbf{r}_q=[x_{\text{r}_q},y_{\text{r}_q}]^T$ being the $q$-th MA location;
and ${\bf z}(t)\in \mathbb{C}^{Q\times 1}$ is the additive Gaussian noise with power $\sigma^2$.
For the $n$-th sampling with interval $T$,
the sampled received signal is expressed as
\begin{equation}\label{EQU-90}
{\bf y}^{\text{me}}[n]={\bf y}^{\text{me}}(nT)=\sum_{l=1}^{L}{\bf h}_l\sqrt{p_t}s(nT-\tau_l)+{\bf z}(nT),
\end{equation}
 where ${\bf z}(nT)$ is the sampled noise, $0\leq n\leq N-1$, with $N$ being the number of samples.
Next, by exploiting the correlation between the received signal ${\bf y}^{\text{me}}[n]$ in (\ref{EQU-90})
and the array response vector ${\bf f}(\theta,\phi)$ in (\ref{array response}), the estimated PAS can be obtained as
\begin{equation}\label{PAS}
\begin{split}
{\rm {PAS}}(\theta,\phi)={\bf f}^H(\theta,\phi)\left(\sum_{n=0}^{N-1}{\bf y}^{\text{me}}[n]({\bf y}^{\text{me}}[n])^H\right){\bf f}(\theta,\phi).
\end{split}
\end{equation}

As such, the estimated number of multipath components $\hat{L}$ and estimated elevation and azimuth AoAs $\left\{{\hat \theta_l}, {\hat \phi_l}\right\}_{l=1}^{\hat L}$ can be obtained by spectrum searching \cite{31266, 10349843}.
Next, per-path zero-forcing (ZF) beamforming  ${\bf w}_l\in \mathbb{C}^{Q \times 1}$ is used to extract the $l$-th path from the received signal \cite{zhou2024single}, which is designed as
\begin{equation}\label{zf_beamforming}
\mathbf{w}_l=\frac{1}{\sqrt{Q}}\left( \mathbf{I}_Q-\mathbf{F}_l\left( \mathbf{F}_{l}^{H}\mathbf{F}_l \right) ^{-1}\mathbf{F}_{l}^{H} \right) \mathbf{f}( \hat{\theta}_l,\hat{\phi}_l ),\enspace 1\leq l \leq \hat L,
\end{equation}
where $\mathbf{F}_l=\left[\mathbf{f}( \hat{\theta}_i,\hat{\phi}_i )\right]_{1\leq i\leq \hat{L}, i\neq l}\in \mathbb{C}^{Q \times (\hat{L}-1)}$. Therefore, the received signal for the $l$-th path is calculated as
\begin{equation}\label{zf_beamforming}
{y}^{\text{me}}_l[n]=\mathbf{w}^H_l\mathbf{y}^{\text{me}}[n].
\end{equation}

It is noted that in the process of utilizing OFDM radar method \cite{2014ofdmradar}, the system response should be calibrated to eliminate the effects of the prototype itself. Therefore, it is necessary to measure the frequency response $H_{\rm {sys}}(f)$ of the developed prototype.
By directly connecting the TX and RX with an RF cable and then transmitting a system response measurement signal, the system frequency response $H_{\rm {sys}}(f)$ can be obtained. Consequently, based on the calibrated channel response, the estimated propagation delay $\{\hat \tau_l\}_{l=1}^{\hat{L}}$ and estimated small-scale amplitude $\{\hat \alpha_l\}_{l=1}^{\hat{L}}$ are thus obtained according to the OFDM radar method.

 Next, in order to obtain the PDS, it is necessary to first perform FFT to the received signal in (\ref{EQU-90}). Thus, the frequency-domain received signal ${\bf Y}_{\text{FFT}} [i,m]\in \mathbb{C}^{Q\times 1}$ for the $m$-th OFDM symbol at the $i$-th sub-carrier is expressed as
 \begin{small}
\begin{equation}\label{FT}
{\bf Y}_{\text{FFT}}[i,m]=\text{FFT}\{{\bf y}^{\text{me}}[n]\}=\sum_{n=0}^{I-1} {\bf y}^{\text{me}}\left[n-\frac{mT_o+T_c}{T}\right]e^{-j2\pi i \frac{n}{I}},
\end{equation}
\end{small}%
where $1\leq m \leq M$, $1\leq i \leq I$.
Then, after system calibration, the frequency response at the $i$-th sub-carrier for $Q$ measurement positions ${\bf H}_{\text{FFT}}[i]\in \mathbb{C}^{Q\times 1}$ is obtained by
\begin{small}
\begin{equation}\label{EQU-9}
{\bf H}_{\text{FFT}}[i]=\sum_{m=0}^{M-1}\frac{{\bf Y}_{\text{FFT}} [i,m]}{\sqrt{p_t}\cdot\text{FFT}\{s[n]\}\cdot H_{\rm {sys}}[i]}=\sum_{m=0}^{M-1}\frac{{\bf Y}_{\text{FFT}} [i,m]}{\sqrt{p_t}b_{i,m}H_{\rm {sys}}[i]},
\end{equation}
\end{small}%
 where $H_{\text{sys}}[i]=H_{\text{sys}}(i\Delta f)$.
Subsequently, by performing IFFT on ${\bf H}_{\text{FFT}}[i]$ in (\ref{EQU-9}), the delay-domain channel for $Q$ measurement positions ${\bf h}[n]\in \mathbb{C}^{Q\times 1}$ is obtained as
\begin{equation}\label{EQU-10}
{\bf h}[n]=\!\mathrm{IFFT}\{{\bf H}_{\text{FFT}}[i]\}=\frac{1}{I}\!\sum_{i=0}^{I-1}{\!}{\bf H}_{\text{FFT}}[i]e^{j2\pi n\frac{i}{I}}, \enspace 0\leq n\leq I-1.
\end{equation}

As a result, the estimated PDS for the $q$-th measurement position ${\rm {PDS}}_q(n\tau_d)$, with $\tau_d=\frac{1}{I\Delta f}$ being the delay step size \cite{7848705}, is given by
\begin{equation}\label{PDS}
{\rm {PDS}}_q(n\tau_d)=\frac{\left|\left[{\bf h}[n]\right]_{q}\right|^2}{\max_n\left\{\left|\left[{\bf h}[n]\right]_{q}\right|^2\right\}}, \enspace 0\leq n\leq I-1.
\end{equation}






\section{Experimental and simulation results}

In this section, we present both the experimental and simulation results of MA communication system at carrier frequencies of 3.5 GHz and 27.5 GHz.
In the experimental measurement, for the 3.5 GHz scenario, the MA moves in a range of about $6\lambda$ with a step size of about $0.01\lambda$. For the 27.5 GHz case, the MA moves in a square area of size $5\lambda \times 5\lambda$ with a step size of around $0.05\lambda$ in both vertical and horizontal directions.
For channel sounding, the movement step sizes are set to be larger than those in power measurement to reduce the computational complexity of channel estimation.
Specifically,
in channel sounding, the MA moves in a square region of  $6\lambda \times 6\lambda$ with a step size of around $0.06\lambda$ in both vertical and horizontal directions at sub-6 GHz, and moves in the same square region as that in performance measurement with a step size of $0.1\lambda$ in both vertical and horizontal directions at mmWave. The PSI for TX1 is estimated based on the channel sounding results. 
Due to the absence of gain calibration data for the PA and LNA inside the USRP, the developed prototype cannot accurately obtain the local-average received power $G$ in \eqref{received_power}.
Therefore, the simulation results consider the small-scale channel power gain in \eqref{received_power} rather than the absolute received power, whose  parameters are set to be the same as the estimated parameters $\left\{\hat \theta_l,\hat\phi_l,\hat \alpha_l,\hat \tau_l\right\}_{l=1}^{\hat{L}}$  for TX1 as in Section \ref{Estimation of Channel Parameters}. Thus, it is calculated according to
\begin{equation}\label{simulation_result}
\begin{aligned}
    g^{\text{sim}}(\mathbf{r}) =  \left|\sum\limits_{l=1}^{\hat{L}}\hat{\alpha}_l\exp\left(-j2\pi\left(\frac{\hat{d}_l(\mathbf{r})}{\lambda}+\hat{\tau}_lf_c\right)\right)\right|^2,  \forall\mathbf{r}\in\mathcal{C}_\text{r},
\end{aligned}
\end{equation}
where $\hat{d}_l(\mathbf{r})=x_\text{r}\cos\hat{\theta}_l\sin\hat{\phi}_l+y_\text{r}\sin\hat{\theta}_l$.

\subsection{Results for 3.5 GHz}
\begin{figure}[htbp]
\centering
\includegraphics[width=1\columnwidth]{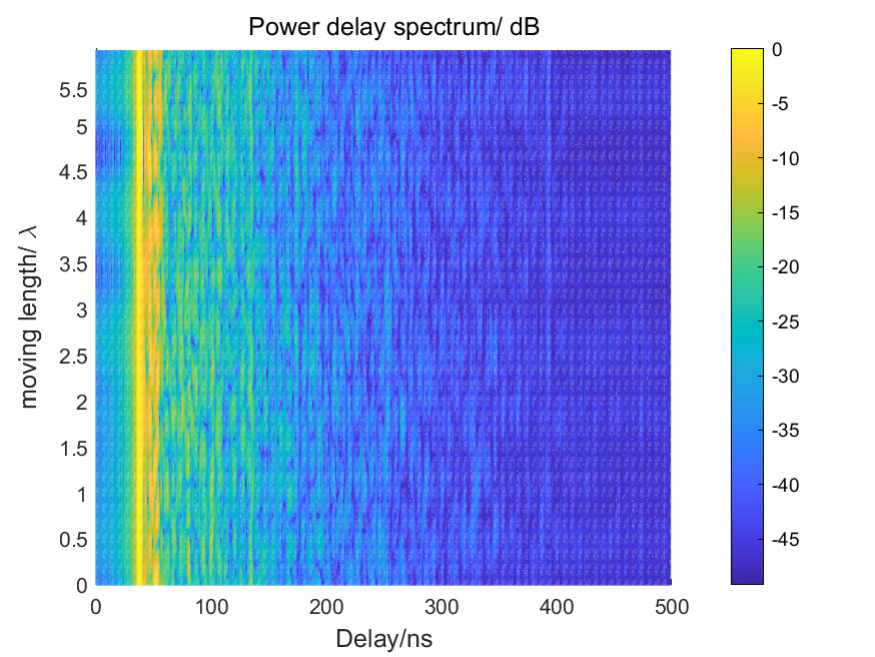}
\caption{The estimated PDS (dB) for TX1 obtained from the channel sounding results, for carrier frequency of 3.5 GHz.}
\label{fig:pds35G}
\end{figure}

Fig. \ref{fig:pds35G} shows the estimated PDS obtained via (\ref{PDS}) for TX1 based on the channel sounding results for 3.5 GHz.
As shown in Fig. \ref{fig:pds35G}, the LoS component remains unchanged within the movement range of the RX MA, which verifies the spatial stationarity characteristic of the channel. This is expected since the size of the moving range is small enough such that the TX is located in the far-field region of the RX. In addition, it is observed from Fig. \ref{fig:pds35G} that the NLoS components vary along the moving range.
These variations are mainly caused by the coherent and non-coherent superposition of adjacent multipath components within a very small time interval in the time domain.

\begin{figure}[htbp]
\centering
\includegraphics[width=1\columnwidth]{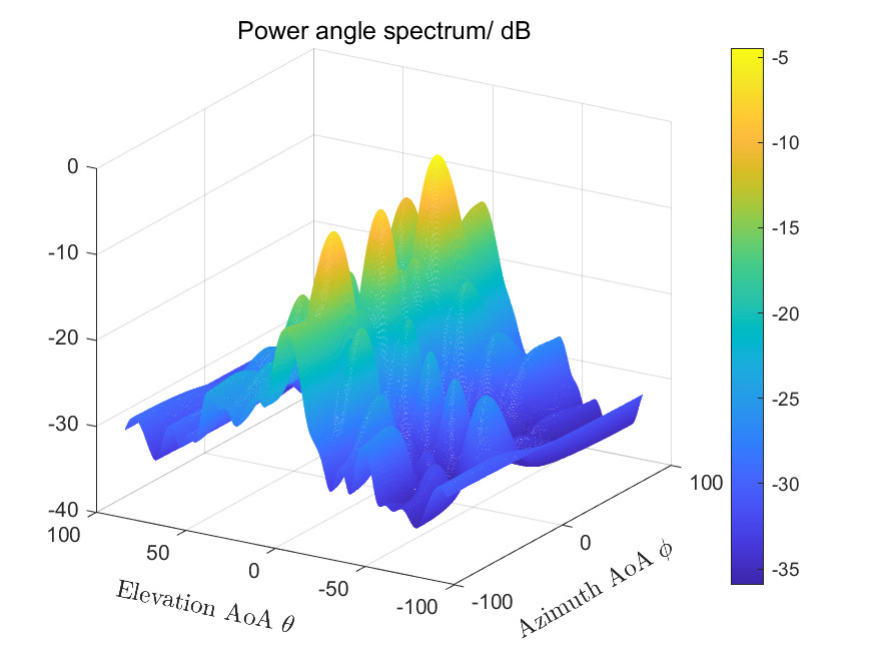}
\caption{The estimated PAS (dB) for TX1 obtained from the channel sounding results at 3.5 GHz.}
\label{fig:PAS35G}
\end{figure}
Fig. \ref{fig:PAS35G} shows the estimated PAS (dB) based on (\ref{PAS}) for TX1 at 3.5 GHz.
As shown in Fig. \ref{fig:PAS35G}, there are multiple significant propagation paths, due to the rich scattering path characteristics of the sub-6 GHz channel.
In Fig. \ref{fig:PAS35G}, $\hat{L}=5$ dominant estimated propagation paths are considered, and the estimated parameters of each path are listed in TABLE \ref{simulation_parameter1}.
As shown in TABLE \ref{simulation_parameter1}, the sum power of the five dominant paths approximately reaches 99.5\% of the total power, i.e., $\sum_{l=1}^{\hat{L}}\hat{\alpha}_l^2\approx0.995$.

\begin{table}[ht]
\begin{center}\caption{The estimated PSI for TX1 at 3.5 GHz.}
\label{simulation_parameter1}
\begin{tabular}{|c|c|c|}
\hline
    carrier frequency    & path index &  parameters of propagation
 path\\ \hline
\multirow{12}{*}{3.5 GHz} & \multirow{2}{*}{$l=1$} & $\hat{\theta}_1=-0.5^{\circ}$, $\hat{\phi}_1=49.5^{\circ}$, \\
                       ~ &  ~                     & $\hat{\alpha}_1=0.6284$, $\hat{\tau}_1=34.8\enspace\mathrm{ns}$      \\ \cline{2-3}
                       ~ & \multirow{2}{*}{$l=2$} & $\hat{\theta}_2=2^{\circ}$, $\hat{\phi}_2=2^{\circ}$,        \\
                       ~ &  ~                     & $\hat{\alpha}_2=0.6075$, $\hat{\tau}_2=22.6\enspace\mathrm{ns}$      \\ \cline{2-3}
                       ~ & \multirow{2}{*}{$l=3$} & $\hat{\theta}_3=1^{\circ}$, $\hat{\phi}_3=-47^{\circ}$,      \\
                       ~ & ~                      & $\hat{\alpha}_3=0.4128$ , $\hat{\tau}_3=34.8\enspace\mathrm{ns}$     \\ \cline{2-3}
                       ~ & \multirow{2}{*}{$l=4$} & $\hat{\theta}_4=15.5^{\circ}$, $\hat{\phi}_4=51.5^{\circ}$,  \\
                       ~ & ~                      & $\hat{\alpha}_4=0.1798$ , $\hat{\tau}_4=36.7\enspace\mathrm{ns}$     \\ \cline{2-3}
                       ~ & \multirow{2}{*}{$l=5$} & $\hat{\theta}_5=14^{\circ}$, $\hat{\phi}_5=-52^{\circ}$,     \\
                       ~ & ~                      & $\hat{\alpha}_5=0.1673$ , $\hat{\tau}_5=40.5\enspace\mathrm{ns}$     \\ \cline{2-3}
                       \hline
\end{tabular}
\end{center}
\end{table}

\begin{figure}[htbp]
\centering
\includegraphics[width=1\columnwidth]{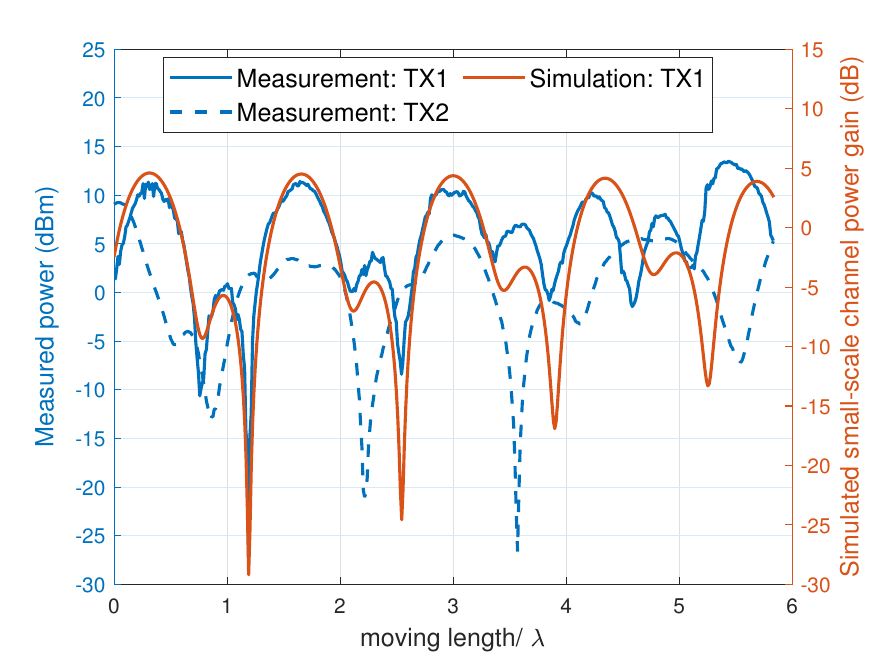}
\caption{The measured power (dBm) for TX1 and TX2 and the simulated small-scale channel power gain (dB) for TX1, for carrier frequency of 3.5 GHz.}
\label{fig:1D_gain}
\end{figure}

Fig. \ref{fig:1D_gain} shows the results of measured power (dBm) in (\ref{EQU-8}) for TX1 and TX2 and the simulated small-scale channel power gain (dB) $g^{\text{sim}}(\vec{r})$ in \eqref{simulation_result} for TX1 at 3.5 GHz, where the simulation results utilize the estimated PSI for TX1 in TABLE \ref{simulation_parameter1}. 
It is observed from Fig. \ref{fig:1D_gain} that the variation range of the measured power for TX1 is over 40 dB in the wavelength range, where the maximum and minimum measured power are 13.3 dBm and -27.1 dBm, respectively. This is expected since multipath components in the wireless channel are coherently and non-coherently superimposed, resulting in channel power gain between the TX and RX varying with the position of RX MA.
By comparing the measurement results with the simulation results, it can be found that both have similar variation trends.
The gap between them is mainly caused by the estimation error of channel parameters.
In addition, it is also observed from Fig. \ref{fig:1D_gain} that a slight variation of TX position results in significant changes in the measured power. This is expected, since change in TX position causes variations in the delay of propagation paths. In this case, even at the same MA position, the measured power will vary due to the coherent and non-coherent  superposition of multipath components.
The above experimental results demonstrate the great potential of MA for improving communication performance.
Furthermore, channel sounding with a relatively large step size and performance measurement with a few measurements can be jointly utilized to guide the MA's movement. Specifically, the MA can first move to an initial rough position according to the simulation results by utilizing the estimated PSI via channel sounding, and then the MA is moved to an accurate position according to a few performance measurements around the initial position with refined movement step size.

\subsection{Results for 27.5 GHz}
\begin{figure}[htbp]
\centering
\includegraphics[width=1\columnwidth]{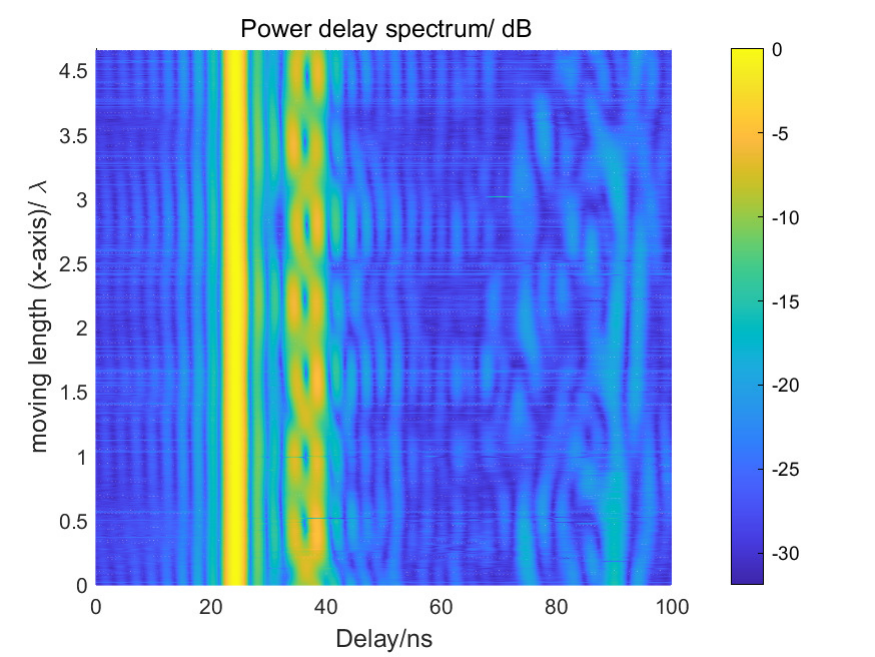}
\caption{The estimated PDS (dB) of the RX MA in the horizontal direction for TX1 obtained from the channel sounding, for carrier frequency of 27.5 GHz.}
\label{fig:cpds28G}
\end{figure}

Fig. \ref{fig:cpds28G} shows the estimated PDS based on (\ref{PDS}) of the RX MA in the horizontal direction for TX1 obtained from the channel sounding, for carrier frequency of 27.5 GHz.
It is observed from Fig. \ref{fig:cpds28G} that there are only three significant propagation paths, which demonstrates the multipath sparsity of mmWave channels.
In addition, the LoS component exhibits the spatial stationarity characteristic,
since TX1 is located in the far-field region of the RX.
Furthermore, it is also observed from Fig. \ref{fig:cpds28G} that the NLoS components vary in the measurement range.
Similar to the results at 3.5 GHz, the variations are also caused by the coherent and non-coherent superposition of adjacent multipath components in the time domain, which have a very small time interval and come from different directions, as indicated by the PAS results below.

\begin{figure}[htbp]
\centering
\includegraphics[width=1\columnwidth]{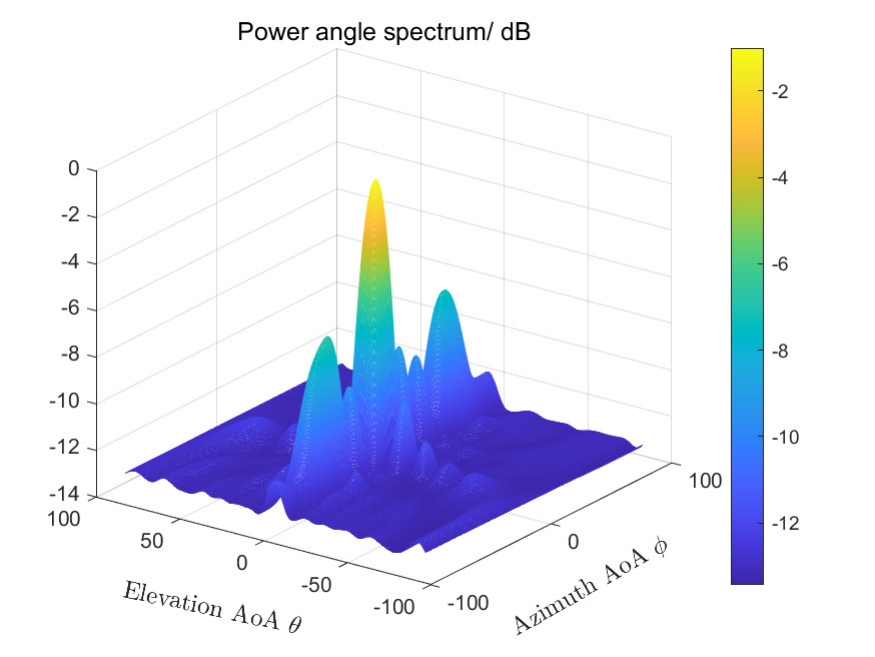}
\caption{The estimated PAS (dB) for TX1 obtained from channel sounding, for carrier frequency of 27.5 GHz.}
\label{fig:PAS28G}
\end{figure}

Fig. \ref{fig:PAS28G} shows the estimated PAS (dB) in (\ref{PAS}) for TX1 at 27.5 GHz.
It is observed that there are three paths with significant power, and the LoS component accounts for a relatively high proportion of the total power, which is consistent with the estimated PDS in Fig. \ref{fig:cpds28G}.
The estimated parameters of the $\hat{L}=3$ dominant propagation paths are given by TABLE \ref{simulation_parameter2}
, and their sum power is approximately equal to the total power of all paths, i.e., $\sum_{l=1}^{\hat{L}}\alpha_l^2\approx1$. In addition, it is found that the angles of the three multipath components at mmWave in TABLE \ref{simulation_parameter2} are approximately the same as those of the first three paths at sub-6 GHz in TABLE \ref{simulation_parameter1}, due to the same propagation environment.

\begin{table}[ht]
\begin{center}\caption{The estimated PSI for TX1 at 27.5 GHz.}
\label{simulation_parameter2}
\begin{tabular}{|c|c|c|}
\hline
    carrier frequency    & path index &  parameters of propagation
 path\\ \hline
 \multirow{6}{*}{27.5 GHz} & \multirow{2}{*}{$l=1$} & $\hat{\theta}_1=3^{\circ}$, $\hat{\phi}_1=2^{\circ}$,       \\
                        ~ &  ~                     & $\hat{\alpha}_1=0.8886$, $\hat{\tau}_1=22.7\enspace\mathrm{ns}$      \\ \cline{2-3}
                        ~ & \multirow{2}{*}{$l=2$} & $\hat{\theta}_2=2.5^{\circ}$, $\hat{\phi}_2=-48.5^{\circ}$,  \\
                        ~ &  ~                     & $\hat{\alpha}_2=0.3423$, $\hat{\tau}_2=35.3\enspace\mathrm{ns}$      \\ \cline{2-3}
                        ~ & \multirow{2}{*}{$l=3$} & $\hat{\theta}_3=2.5^{\circ}$, $\hat{\phi}_3=49.5^{\circ}$,   \\
                        ~ & ~                      & $\hat{\alpha}_3=0.3053$, $\hat{\tau}_3=34.8\enspace\mathrm{ns}$      \\ \hline
\end{tabular}
\end{center}
\end{table}

Fig. \ref{fig:7} shows the simulated small-scale channel power gain (dB) for TX1 and measured power (dBm) for TX1 and TX2, respectively, at carrier frequency of 27.5 GHz.
The simulation result utilizes the estimated PSI shown in TABLE \ref{simulation_parameter2}.
As illustrated in Fig. \ref{fig:7} (a), the variation range of simulated channel power gain is over 17 dB, where the maximum and minimum channel gains are 3.8 dB and -13.3 dB, respectively. This is due to the coherent and non-coherent superposition of multipath components.
In Fig. \ref{fig:7} (b), the variation range of the measured power for TX1 is over 23 dB, where the position of the maximum value in the measurement results matches that in the simulation results well.
 The minor difference between measurement and simulation results may be caused by the angle estimation error of propagation paths and the imperfect radiation pattern of mmWave receiving antenna.
In addition, by observing the measurement results for TX at different positions in Fig. \ref{fig:7} (b) and  Fig. \ref{fig:7} (c), it is found that the movement of TX position causes significant variation of measured power in the measured region, which is due to the change in propagation path delays.
The experimental results demonstrate that the MA technology can significantly improve the received signal power.
Furthermore, the MA can be guided to move to a relatively accurate position according to simulation results, which are  obtained from accurately estimated channel parameters thanks to the multipath sparsity of the mmWave channel.
Then, the MA can be moved to a more accurate position by performing a few additional performance measurements with a refined movement step size.

\begin{figure}[htbp]
\centering
\subfigure[Simulation results for TX1]{
	\includegraphics[width=1\columnwidth]{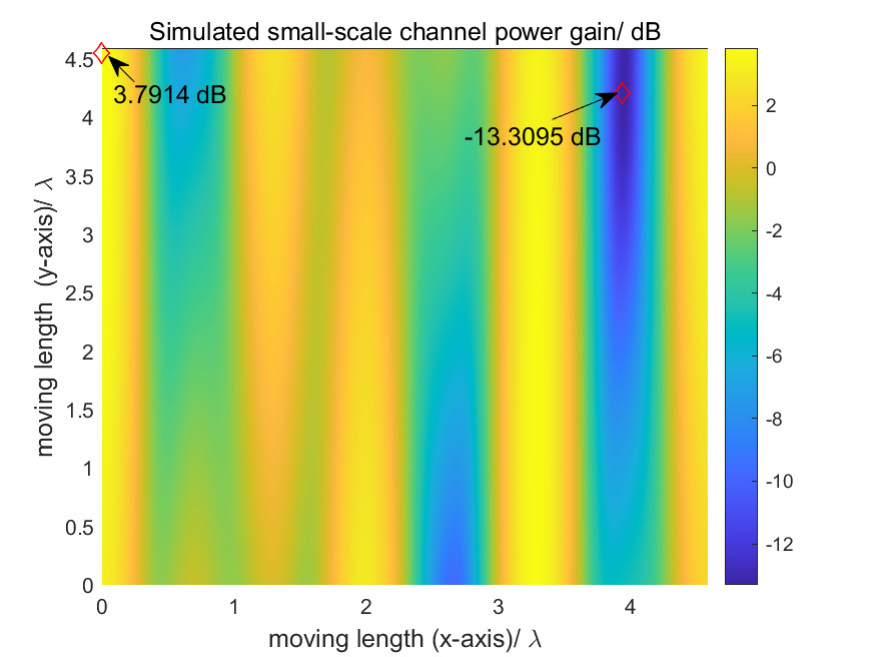}}
\subfigure[Measurement results for TX1]{
	\includegraphics[width=1\columnwidth]{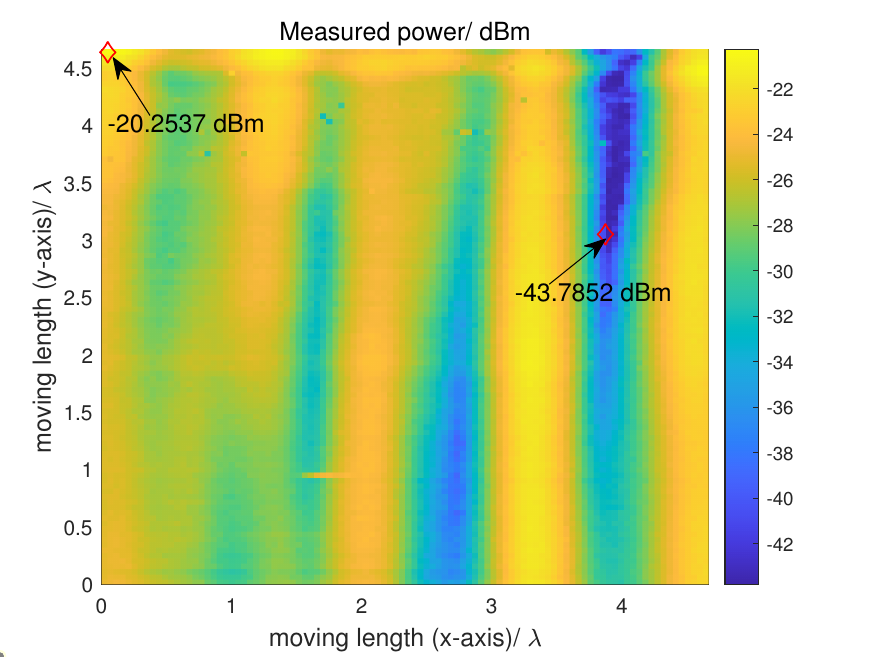}}
 \subfigure[Measurement results for TX2]{
	\includegraphics[width=1\columnwidth]{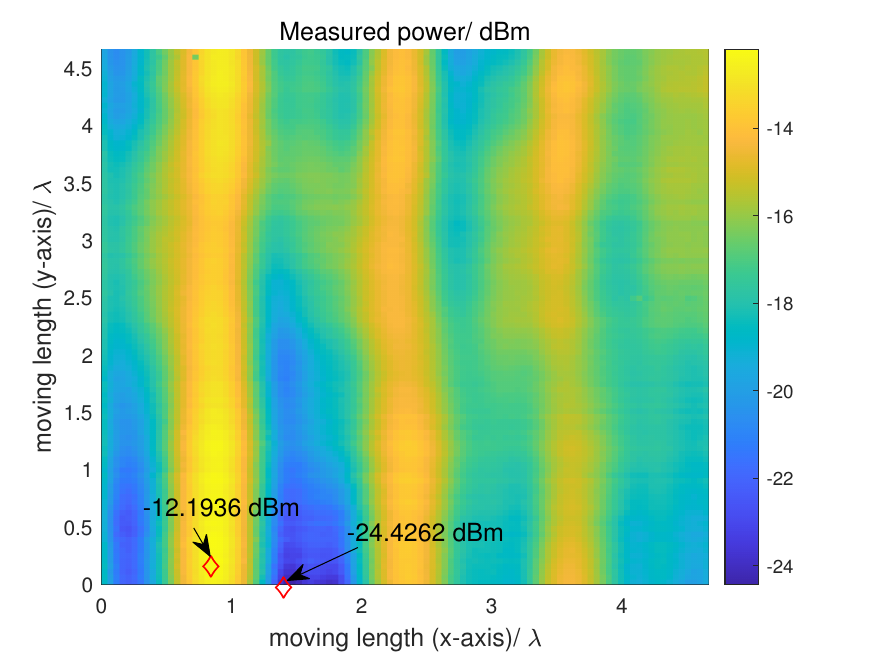}}
\caption{Simulated small-scale channel power gain (dB) for TX1 and measured power (dBm) for TX1 and TX2, for carrier frequency of 27.5 GHz.}
\label{fig:7}
\end{figure}

\section{conclusion}

In this paper, a prototype for MA wireless communication was built to practically demonstrate the benefit of MA technology in improving communication performance.
By obtaining the PSI between the TX and the receiving region, the MA system can achieve spatial diversity gain by flexibly varying antenna positions and fully utilizing channel spatial variations.
We first built the MA communication experimental prototype with highly accurate movement control, by developing both a hardware platform and a software program.
Then, the MA system prototype was tested at 3.5 GHz and 27.5 GHz.
The experimental results were compared with the simulation results, demonstrating the great performance gain of MA via antenna movement within the wavelength range in practical environments.

\begin{appendices}


\end{appendices}
\bibliographystyle{IEEEtran}
\bibliography{./header_short,./bibliography1}

\end{document}